\documentclass[sigconf]{acmart}

\usepackage{booktabs}
\usepackage{tabularx}
\usepackage{array}
\usepackage{multirow}
\usepackage{graphicx}
\usepackage{subcaption}
\usepackage{balance}
\usepackage{listings}
\setcopyright{none}
\renewcommand\footnotetextcopyrightpermission[1]{}

\acmConference[CIKM '26]{The 35th ACM International Conference on Information and Knowledge Management}{November 7--11, 2026}{Rome, Italy}
\acmYear{2026}
\copyrightyear{2026}
\acmDOI{}
\acmISBN{}

\title{Role of  Personality in Conversational Information Seeking}

\author{Abdisalam Abukar}
\affiliation{%
\institution{GAIR-Lab - School of Computing Science,}
  \institution{University of Glasgow}
  \country{United Kingdom}
}
\email{2618966A@student.gla.ac.uk}

\author{Junchen Fu}
\authornote{Corresponding Author.}
\affiliation{%
\institution{GAIR-Lab - School of Computing Science,}
  \institution{University of Glasgow}
  \country{United Kingdom}
}
\email{j.fu.3@research.gla.ac.uk}

\author{Chengli Zhai}
\affiliation{%
\institution{School of Education,}
  \institution{University of Glasgow}
  \country{United Kingdom}
}
\email{c.zhai.1@research.gla.ac.uk}

\author{Joemon M. Jose}\affiliation{
\institution{GAIR-Lab - School of Computing Science,}
\institution{University of Glasgow}\streetaddress{}\city{Glasgow}\country{United Kingdom}}
\email{joemon.jose@glasgow.ac.uk}

\begin{document}

\begin{abstract}
Large language models (LLMs) are increasingly used for information seeking, where users find, compare, and evaluate information through dialogue. In this role, the assistant (LLM) does more than retrieve or generate content — it actively mediates how users articulate constraints, pose follow-up questions, verify claims, and determine when an answer is sufficient for action.  Despite this influence, little is understood about how user personality, assistant personality, and task context jointly shape such interactions. We examine  personality as a controllable variable in conversational information seeking, with the aim of understanding its effects on user behaviour and interaction quality.
Hence, we designed a controlled, within-subject interactive study in which assistant personality and task type were experimentally varied, while participant personality was measured using Big Five scores. Twenty-six participants each completed three information-seeking tasks across three assistant personality conditions — an extraverted style, a conscientious style, and a neutral baseline. The tasks spanned three distinct domains: exploratory travel planning, comparative smartphone shopping, and verification-sensitive health and diet information seeking. Data were collected through conversation logs, behavioural traces, post-interaction questionnaires, an exit questionnaire, and Big Five personality measures.
The assistant conditions were behaviourally distinguishable: the extraverted assistant produced longer assistant turns, the conscientious assistant yielded higher user word share and more turns, and the neutral baseline fell between them. The strongest effect was a task-by-assistant interaction on trust and delegation, with preferred styles varying by task. 
 No global winner emerged, but participants strongly favoured style choice or adaptation. These findings position assistant personality as a context-sensitive, interactional design variable rather than a globally optimisable system property.
\end{abstract}



\maketitle

\section{Introduction}

Information seeking is increasingly mediated through dialogue~\cite{zamani2023conversational, volkel22,  BatistaJJ26}. Instead of formulating one query, scanning a ranked list, and navigating documents, users now ask conversational agents for recommendations, explanations, comparisons, and summaries~\cite{dalton22,fu2024iisan,fu2024exploring}. This shift is visible in large-scale analyses of consumer chatbot usage, where practical guidance, information seeking, and writing form major categories of interaction \cite{chatterji2025howpeopleusechatgpt}. The adoption of chatbot-based interfaces in information retrieval marks a transition from one-shot querying to continuous, natural language interaction~\cite{BatistaJJ26, potiagalova2025comparative}. Users now engage in dialogue with the system—incrementally refining constraints, probing assumptions, and requesting validation—until results are sufficient for action. In this interactive setting, retrieval effectiveness is shaped not only by relevance but also by how the system communicates, motivating investigation into the role of personality in conversational search \cite{Byeon26, han2026personalityexpressioncontextslinguistic, jayasiriwardene2026fixedflexibleshapingai, zhang2025decodinghumanllmcollaborationcoding}.

This shift makes conversational style a consequential design factor in conversational information retrieval. In chatbot-based search and question answering, an assistant may engage in more mixed‑initiative behavior by asking clarifying questions \cite{Aliannejadi19}, provide more directive recommendations, impose stronger dialogue structure \cite{Liang24}, or adopt a more socially expressive tone \cite{Galland22}. Such stylistic choices influence not only the flow of the interaction but also user trust, engagement, and interaction efficiency, even when the underlying retrieval or language model and task remain unchanged. Recent large language models (LLMs) make these variations easy to instantiate through prompting: the same model can be instructed to respond in a more extraverted \cite{volkel22}, cautious, structured, or neutral manner. While these styles do not correspond to personality in the psychological sense, they function as an expressed assistant personality that users perceive and adapt to during conversational search.

At the same time, conversational information seeking is both task-dependent and user-dependent. Different tasks place different demands on interaction: exploratory scenarios such as travel planning may benefit from richer dialogue, proactive clarification, and socially engaging responses, whereas goal-oriented tasks such as shopping or troubleshooting often prioritize efficiency, decisiveness, and low interaction cost. At the same time, users vary in their conversational preferences, tolerance for verbosity, and expectations of system initiative. As a result, the same assistant personality may foster trust and engagement for one user while reducing efficiency or satisfaction for another \cite{cai2024communication}. Personality-aware conversational systems therefore introduce a two-sided personalization problem in conversational IR: the assistant’s expressed style must be aligned with both the task context and the user’s interaction preferences in order to support effective, trustworthy, and efficient information seeking.

We study this problem through a controlled user study in which participants interacted with three LLM assistants across three information-seeking tasks. The assistants were implemented using the same base model (gpt 4.1) but different system prompts. Two prompts expressed Big Five-related styles: Extraversion and Conscientiousness. The third prompt was a neutral baseline. The tasks were selected to capture three common information-seeking contexts: open-ended exploration, comparative decision making, and verification-sensitive advice. The study combines behavioural traces, questionnaires, and participant personality measures to examine how assistant style, participant personality, and task type jointly shape conversational information seeking.

The paper is organized around the following research questions:

\noindent
\textbf{RQ1.} How does user personality shape conversational information‑ seeking behaviour when interacting with an LLM assistant?\\
\textbf{RQ2.} How does the effectiveness of different assistant personality expressions depend on user personality?\\
\textbf{RQ3.} How does the effectiveness of assistant personality expressions differ across information‑seeking task types?

The paper's contribution is empirical rather than prescriptive. We do not advocate for a single optimal assistant personality. Instead, we provide evidence that: (i) assistant style produces distinguishable behavioural interaction patterns; (ii) task context moderates trust and delegation outcomes; and (iii) participant–assistant personality compatibility explains more variance than broad trait-to-behaviour effects. These results support a contextually grounded view of personality-aware LLM design, in which style is treated as a situational interaction variable rather than a universally beneficial system property. In this sense, assistant style is not a cosmetic property but a functional one: it shapes how initiative is distributed, how user constraints are surfaced, how strongly recommendations are framed, and how claims are evidentially grounded.

\section{Related Work}

\subsection{Conversational Information Seeking}

Conversational information seeking (CIS) treats search as an extended interaction rather than a one-shot query-response event \cite{Aliannejadi19,Liang24, Galland22, radlinski2017framework}. Radlinski and Craswell define conversational search around an action space in which systems can accept natural language requests, ask clarifying questions, present results, incorporate feedback, and accumulate context over time \cite{radlinski2017framework}. Zamani et al. similarly frame CIS as a sequence of interactions in which users and systems jointly develop an information need through dialogue \cite{zamani2023conversational}. This perspective is especially relevant for LLM-based assistants because the user's query often evolves across turns as the system suggests options, requests missing details, or reformulates the task. However, effect of user personality in such interactions are not explicitly studied yet.

Classical conversation theory provides a principled account of why conversational search requires ongoing refinement and repair. According to grounding theory, participants collaboratively establish and maintain shared understanding as interaction unfolds \cite{clark1991grounding}. In conversational IR, this process appears in behaviors such as constraint confirmation, intent clarification, and reference repair. Empirical findings from the TREC Conversational Assistance Track \cite{Daltontrec19, dalton2020castoverview} reinforce this view: manually reformulated queries frequently outperform unrevised conversational inputs, suggesting that contextual understanding and reformulation are fundamental retrieval challenges, not merely interface-level artifacts \cite{dalton2020castoverview}.

LLMs intensify this interactional character because they can generate fluent answers that combine retrieval-like, explanatory, and recommendation-like behaviours in the same interface. However, fluency also creates risk. Hallucination remains a known failure mode of neural generation and LLM systems \cite{ji2023survey}, and retrieval-augmented generation is often proposed as a way to ground generated responses in external evidence \cite{lewis2020retrieval,gao2023ragsurvey}. In user-facing information seeking, this means that trust, verification, and willingness to delegate are not secondary usability properties. They are central outcomes, especially when the task involves consequential or health-related information.   Generative retrieval and recommendation have also emerged as active research directions \cite{sun2026zerogr,fu2026differentiable}. Multimodal foundation models have also been increasingly explored for recommendation \cite{fu2024iisan,fu2025efficient,fu2025crossan}.

\subsection{Assistant Personality and Social Response to Machines}

The idea that assistant style can influence user behaviour is consistent with long-standing work on social responses to technology \cite{cai2024communication}.   Literature argues that people often apply social expectations to computers and media, even when they know that the system is not human \cite{reeves1996mediaequation,nass2000machines}. For conversational systems, this matters because small cues in tone, initiative, politeness, confidence, and structure can make a system appear socially warmer, more methodical, more assertive, or more passive.

Personality psychology offers a vocabulary for describing such differences. The Big Five framework captures broad individual-difference dimensions such as Extraversion and Conscientiousness \cite{goldberg1990bigfive,john1999bigfive}. In an LLM assistant, these dimensions are not measured as stable internal traits. Rather, they are operationalised as expressed interaction styles. Extraversion can be expressed through energy, social warmth, assertiveness, and proactive engagement \cite{costa1995domains,mairesse2008trainable}. Conscientiousness can be expressed through structure, planning, carefulness, order, and methodical presentation \cite{costa1995domains,mairesse2008trainable}. Prior work has investigated personality-like behaviour in language systems and has shown that prompting can affect generated personality impressions \cite{serapiogarcia2023personalityllms,imasaka2024personalityquery}. This makes prompt-based personality manipulation practical for a controlled study because the base model can remain fixed while the system instruction varies. 

\subsection{Personality, Search Behaviour, and Personalization}

Personality has also been linked to information behaviour \cite{Byeon26}. Prior studies suggest that personality traits can be associated with differences in search strategies, information-seeking preferences, and online behaviour \cite{schmidt2016personality,AlSamarraieEldenfriaDawoud2017PersonalityInfoSeeking}. In conversational settings, these differences may appear in how users formulate goals, disclose constraints, ask questions, or verify outputs. However, the relationship is unlikely to be purely direct. A user's personality may not simply predict more or fewer turns. Instead, it may interact with the assistant's style and the task context.

Prior work on style and alignment in information-seeking conversation is especially relevant here. It shows that interaction style can be measured, that users may align to an agent's conversational style, and that mismatch can affect perceived effort and smoothness \cite{thomas2018stylealignment,thomas2020expressionsstyle}. Task context also matters: task-based information searching argues that search behaviour should be interpreted in relation to the underlying task, exploratory search differs from narrower lookup or comparison, and online health information often places special weight on trustworthiness and evidence \cite{kumpulainen2017task,marchionini2006exploratory,sun2019consumer}. These strands motivate a compatibility view in which assistant style is not only a global preference, but a possible fit between the user, the assistant, and the information-seeking ecology.

Trust in a conversational assistant is partly a function of fit: when an assistant's expressed style aligns with a participant's own communicative preferences, the interaction tends to feel more coherent and reliable. Conversely, a mismatch may introduce friction — an extraverted assistant may feel engaging to one user but verbose to another, while a methodical assistant may feel reassuring in one task but obstructive in another. The present study sits at the intersection of conversational search and personality-aware interaction. From conversational search, it takes the view that user behaviour unfolds across a sequence of turns rather than a single query. From personality-aware interaction, it takes the view that expressed style can alter how a system is experienced even when the underlying task remains constant. The empirical question is whether these two perspectives converge in measurable ways: whether style shifts observable search behaviour, whether participant traits moderate those shifts, and whether different information-seeking tasks reward different assistant styles.

\section{Method}

\subsection{Study Design}
To answer our research questions, we have devised a user study. Three working hypotheses guided the study. H1 predicted that participant personality would systematically predict conversational behaviour (e.g. message length, turn count, question frequency). H2 predicted that assistant effects would be moderated by participant–assistant compatibility, with better-reported   where assistant style matched the participant's own Extraversion–Conscientiousness profile. H3 predicted that assistant personality effects would vary by task type, given that exploratory, comparative, and verification-sensitive tasks impose different conversational demands \cite{han2026personalityexpressioncontextslinguistic, jayasiriwardene2026fixedflexibleshapingai}. These were treated as directional expectations rather than claims that every measure would reach significance.

Each participant completed three task-phase interactions. Across these interactions, each participant encountered each assistant condition once and each task once; task–assistant pairings and order were counterbalanced across participants \cite{Kelly09, WhiteRJ02}. 
This was particularly important for RQ2, which examines whether assistant personality effects are moderated by participant personality.

To prevent demand effects, the three assistant conditions were presented under neutral planet names — Titan (Extraversion), Europa (Conscientiousness), and Neptune (neutral baseline) — with the personality mapping disclosed only at the end of the experiments. Task and assistant order were counterbalanced using a Graeco-Latin-square scheme to prevent assistant–task pairings and sequence positions from confounding results \cite{Kelly09}. Each interaction began as a fresh conversation instance (Fig. 1).

\begin{table}[t]
\centering
\caption{Assistant conditions used in the within-subject study.}
\label{tab:conditions}
\small
\begin{tabularx}{\linewidth}{p{0.19\linewidth}p{0.26\linewidth}X}
\toprule
\textbf{Name shown} & \textbf{Condition} & \textbf{Intended interaction style} \\
\midrule
Titan & Extraversion & Energetic, socially warm, proactive, and momentum-building. \\
Europa & Conscientiousness & Structured, methodical, careful, and task-focused. \\
Neptune & Neutral baseline & Minimally stylised, concise, and used as a comparison condition. \\
\bottomrule
\end{tabularx}
\end{table}

\subsection{Task Design}

The tasks were selected to represent contrasting information-seeking ecologies while remaining familiar enough for participants to engage naturally. Table~\ref{tab:tasks} gives the concrete scenario for each task, since the paper's argument depends on the difference between exploratory, comparative, and verification-sensitive information seeking \cite{Orland25, 2009White}. The travel task represented preference-building exploration, the smartphone task represented comparative narrowing, and the health/diet task represented a setting in which credibility and caution were expected to matter more strongly.

\begin{table}[t]
\centering
\caption{Information-seeking tasks used in the study.}
\label{tab:tasks}
\footnotesize
\begin{tabularx}{\linewidth}{>{\raggedright\arraybackslash}p{0.20\linewidth}>{\raggedright\arraybackslash}p{0.40\linewidth}X}
\toprule
\textbf{Task} & \textbf{Scenario and goal} & \textbf{Task ecology} \\
\midrule
Travel planning & Plan a six-day summer trip to Turkey with a \pounds 1000 non-flight budget, mid-range comfort, relaxing activities, and one adventurous element. & Exploratory preference disclosure, option exploration, and iterative refinement. \\
Smartphone shopping & Choose a UK smartphone within \pounds 600--\pounds 800, prioritising camera quality, battery life, long-term value, and openness to Android despite current iPhone use. & Comparative trade-off reasoning, narrowing between alternatives, and convergence toward a final choice. \\
Health/diet information & Evaluate a three-day detox/cleanse claim, including what is true or misleading, risks, safer alternatives, and when to seek help. & Verification-sensitive questioning, evidence seeking, and greater scrutiny of the assistant's guidance. \\
\bottomrule
\end{tabularx}
\end{table}

The use of multiple tasks was deliberate. A single task would have made the design simpler, but it would also have made it difficult to determine whether a style effect was specific to that task. Since conversational information seeking includes exploration, comparison, and verification, the study needed task diversity to test whether assistant personality effects generalized or varied by context \cite{zhang2025decodinghumanllmcollaborationcoding}.

\subsection{Assistant Conditions}

All assistant conditions (Table 1) used the same base LLM and the same general conversation policy. The manipulated component was the system prompt defining the assistant's interaction style. The Extraversion prompt emphasized sociability, energy, engagement, and forward conversational momentum. The Conscientiousness prompt emphasized structure, methodical reasoning, careful organization, and task focus. The neutral prompt avoided a strong personality framing and acted as a baseline.

The prompt-based approach was selected instead of fine-tuning because it allowed the study to hold the model constant and vary only the system instruction. This made the manipulation transparent, lightweight, and replicable. Prompt validation was conducted before the live study by generating assistant outputs and assessing whether the intended trait emphasis was visible in the resulting style. The validation was used to check that the conditions were plausibly distinguishable before deploying them to participants.

The design deliberately focused on Extraversion and Conscientiousness rather than all five Big Five dimensions. These two traits were selected because they are comparatively visible in short text-based interaction and could be translated into clear conversational behaviours without changing the task content. Extraversion was expected to appear through warmth, initiative, and active follow-up. Conscientiousness was expected to appear through structured answers, careful sequencing, and methodical checking. The neutral condition provided a baseline against which these more stylised behaviours could be interpreted.

Early pilot iterations also showed that overly complete assistant responses could turn the task into one-shot answer delivery, reducing opportunities for clarification, correction, and preference disclosure. The shared conversation policy therefore constrained early turns to be concise and limited the number of direct questions. This made personality differences more observable through turn-taking, initiative, pacing, and structure rather than only through long monologic answers.

\subsection{Participants and Procedure}

Twenty-eight participants completed the live study. After cleaning the export, two participant cases were excluded because linked data were missing, leaving a complete-case dataset of 26 participants. Since each participant completed three task phases, the final behavioural dataset contained 78 usable task-phase conversations. Cleaning also removed development records, duplicate or retry records, and incomplete cases that did not correspond to a full participant trajectory. The same 26-participant subset was used for the post-interaction analysis, the exit questionnaire analysis, and the compatibility analysis.

The complete-case sample was young, male-skewed, and largely computing-oriented. Twenty participants were aged 18--24 (76.9\%) and six were aged 25--35 (23.1\%). Twenty-one participants identified as male (80.8\%) and five as female (19.2\%). Most participants reported a student or PGR status, and many reported study or work areas related to computer science, computing, IT, software engineering, or technology. Educational backgrounds ranged from secondary school and College/HND to undergraduate and postgraduate study, with undergraduate and secondary-school qualifications being the most common. Entry-questionnaire responses also indicated substantial prior exposure to AI assistants: most participants had used an AI assistant before, and reported use was mainly daily or weekly. Overall, the sample should therefore be interpreted as a young, male-skewed, computing-oriented, and AI-exposed convenience sample rather than a demographically balanced population sample.

The procedure had four stages. First, participants completed pre-study materials, including consent, an entry questionnaire, and a Big Five personality questionnaire. Second, participants completed three live task phases. In each phase, the participant received the task handout, interacted with the assigned assistant, and then immediately completed a post-interaction questionnaire. Third, after all task phases, participants completed an exit questionnaire comparing the systems. Finally, participants saw the personality reveal. This ordering preserved task-specific impressions during the post-interaction ratings while still allowing comparative judgements after the full set of interactions.

\subsection{Measures and Analysis}

The study combined multiple data sources: conversation logs, message logs, derived conversation statistics, post-interaction questionnaires, exit questionnaires, and Big Five personality scores. The final analytic subsets are summarized in Table~\ref{tab:data}. Conversation statistics included assistant response length, user word share, turn count, user message length, follow-up turns, user question count, and related behavioural trace features. Questionnaire analysis focused on post-interaction constructs including trust and delegation, interaction quality, and an overall post-interaction composite. Open-text comments were reviewed using lightweight thematic coding focused on recurring perceptions of task--style fit, trust, conversational pacing, and interaction friction. Comments could receive multiple codes where relevant. Because the qualitative material consisted of brief post-interaction comments rather than interviews, the coding was used interpretively to contextualize behavioural and questionnaire findings rather than to establish standalone qualitative claims or a full standalone thematic analysis.

\begin{table}[t]
\centering
\caption{Raw exports and final analytic subsets.}
\label{tab:data}
\small
\begin{tabular}{llr}
\toprule
\textbf{Category} & \textbf{Item} & \textbf{Count} \\
\midrule
Raw files & Conversations & 155 \\
 & Messages & 786 \\
 & Questionnaires & 140 \\
 & Conversation statistics & 151 \\
 & BFI score rows & 30 \\
\midrule
Final subsets & Complete participants & 26 \\
 & Usable task-phase conversations & 78 \\
 & Post-interaction subset & 26 \\
 & Exit questionnaire subset & 26 \\
 & BFI-linked compatibility subset & 26 \\
\bottomrule
\end{tabular}
\end{table}

For RQ1, participant Extraversion and Conscientiousness were analysed as predictors of behavioural trace measures. For RQ2, compatibility was operationalised as the participant's trait balance, $z(\mathrm{Extraversion}) - z(\mathrm{Conscientiousness})$, using the two trait-linked assistant conditions. Positive values indicated a more Extraversion-weighted profile and negative values indicated a more Conscientiousness-weighted profile. Neptune was treated as the neutral comparison and excluded from the Titan-Europa compatibility slope. For RQ3, the main analysis examined whether assistant effects differed by task, especially for trust and delegation and the overall post-interaction composite.

The behavioural traces were grouped into three broad constructs. First, participation structure captured who occupied more of the conversation, using assistant response length and user word share. Second, interaction pacing captured how much turn-by-turn exchange occurred, using turn count and follow-up turns. Third, questioning behaviour captured how often participants asked explicit questions or used the assistant to check details. These traces were not interpreted as independent measures of success; rather, they were used to describe how the interaction unfolded under different assistant styles.

Interaction friction was therefore treated as a practical pattern, not as a single number. Repeated repair, excessive clarification, checking, or slow convergence could indicate friction, but longer conversations were not automatically treated as worse. In exploratory tasks, additional turns can indicate engaged preference development; in comparative or verification-sensitive tasks, similar trace values may instead reflect uncertainty or resistance. Trace measures were consequently interpreted jointly with questionnaire scores and participant comments. Table~\ref{tab:measures} summarizes the data sources used in the analysis and the role each source played in interpreting the quantitative and qualitative evidence.

\begin{table}[t]
\centering
\caption{Main data sources and their role in the analysis.}
\label{tab:measures}
\footnotesize
\begin{tabularx}{\linewidth}{>{\raggedright\arraybackslash}p{0.23\linewidth}X}
\toprule
\textbf{Source} & \textbf{Use in analysis} \\
\midrule
Logs & Response length, user word share, turn count, follow-up turns, and question behaviour. \\
Post survey & Task-local evaluations, especially trust/delegation and the overall post-interaction composite. \\
Exit survey & Comparative preferences, personality reveal responses, and attitudes toward choosing or adapting assistant style. \\
BFI scores & Participant Extraversion, Conscientiousness, and Titan-Europa compatibility score. \\
Comments & Qualitative support for interpreting why styles fit or failed in particular tasks. \\
\bottomrule
\end{tabularx}
\end{table}

The interpretation of inferential tests was conservative. Where effects did not survive correction or remained close to conventional significance thresholds, they are reported as directional rather than confirmatory. This is important because the study was designed as a controlled exploratory user study rather than as a large-scale population estimate. The goal was to identify whether assistant personality, participant personality, and task type generated coherent patterns that would justify larger follow-up experiments.

\section{Experiment Setup}

\subsection{System Implementation}
\begin{figure*}[t]
    \centering
    \includegraphics[width=0.9\textwidth]{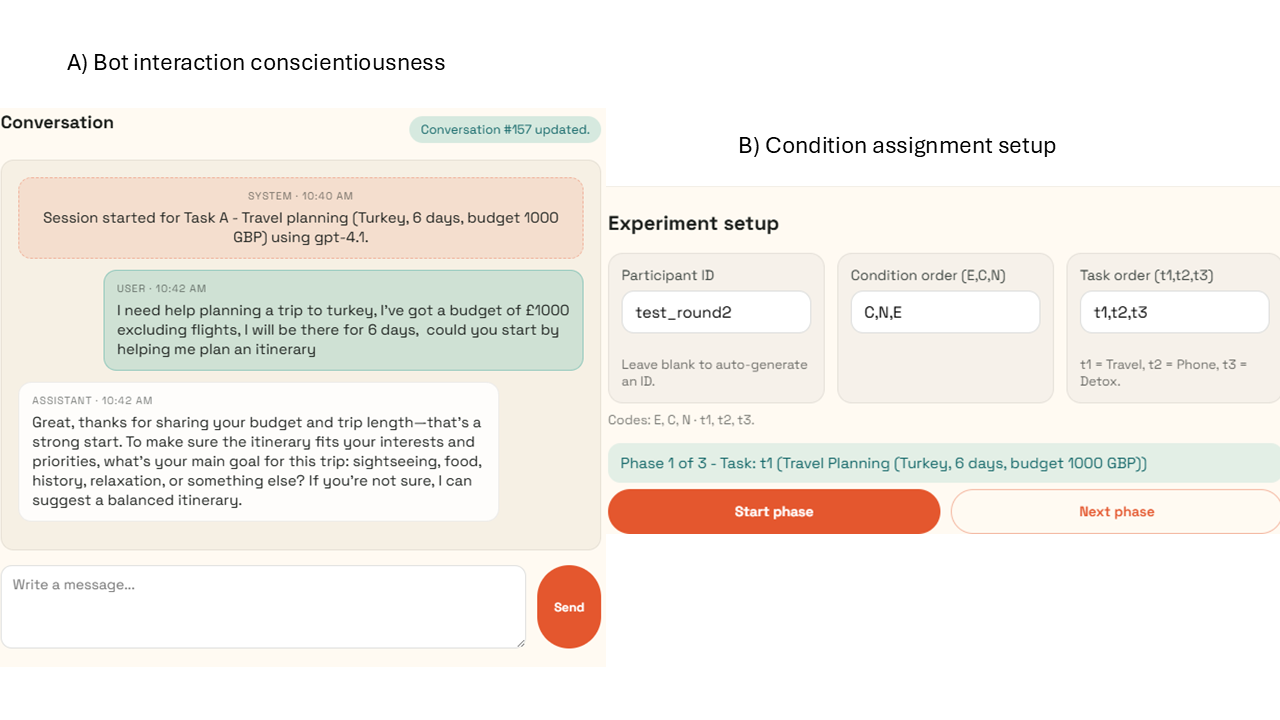}
    \vspace{-0.2in}
    \caption{Custom interface used during the live study. Panel A shows an example initial interaction with the conscientious assistant condition. Panel B shows the experiment setup view, including participant ID, assistant condition order, and task order used for the counterbalanced Graeco-Latin-square assignment.}
    \label{fig:interface-overview}
\end{figure*}

The study was implemented as a custom browser-based conversational interface backed by a server-side application (Figure 1). Pre-study materials and questionnaires were administered externally, while the live interaction took place in the custom system. The system stored conversation-level records, message-level records, questionnaire responses, derived statistics, and BFI scores for later analysis. The task instructions were provided as printed handouts rather than embedded in the assistant prompt. This separated the participant's task context from the assistant's personality manipulation.

During the live session, each assistant used the same underlying model and the same general constraints. The model used in the implementation was \texttt{gpt-4.1}. The key experimental control was that assistant style changed through the prompt while the base model remained fixed \cite{shah25}. This allowed the comparison to focus on expressed assistant personality rather than model capability differences. The system recorded each task phase as a separate conversation, which prevented context from one assistant or task carrying over into the next live phase.

Participants interacted with assistants named Titan, Europa, and Neptune (Figure 1). These names were intentionally neutral and did not disclose the personality condition. The design avoided labels such as ``extraverted assistant'' or ``conscientious assistant'' during the live phases, because such labels could prime expectations and contaminate questionnaire responses. The reveal was delayed until the exit stage so that participants' post-interaction ratings were based on their immediate experience of the interaction rather than on explicit knowledge of the manipulation.

The post-interaction questionnaire was completed immediately after each task phase. 
 The post-interaction questionnaire therefore captured local, task-specific evaluations, including trust, willingness to rely on the assistant, interaction smoothness, and perceived task support.

The exit questionnaire captured comparative judgements that could only be made after participants had used all three systems. It also measured whether the personality reveal changed preferences and whether participants liked the idea of choosing or automatically adapting assistant style. The behavioural trace features were derived from the logs and used to examine how the interaction unfolded independently of questionnaire ratings.

\subsection{Manipulation Validation}

\begin{figure}[t]
    \centering
    \includegraphics[width=\linewidth]{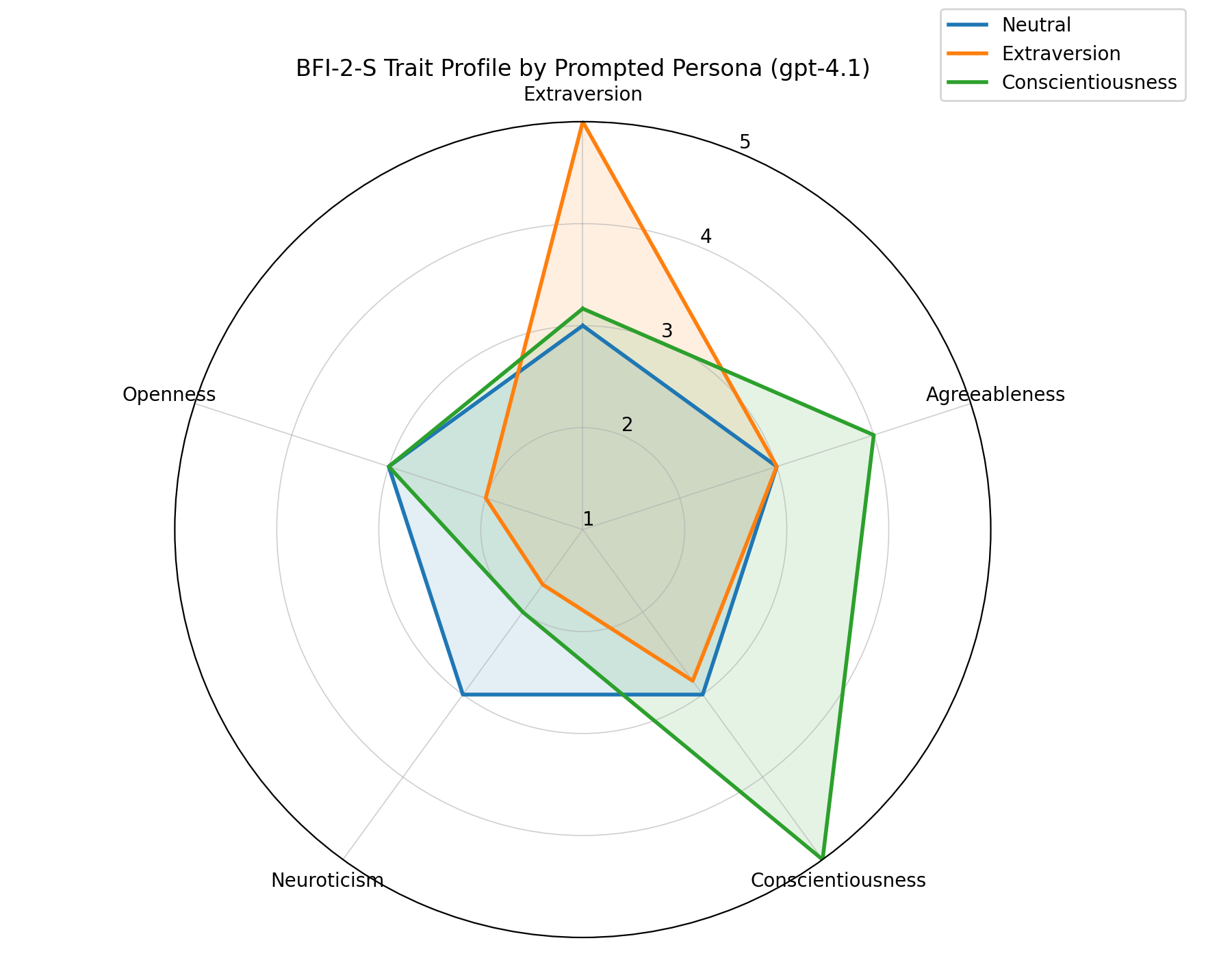}
    \caption{Prompt validation profiles for the three assistant conditions. The plot summarizes the BFI-2-S trait profile generated under each prompt condition and was used before the live study to check that the prompt-defined styles produced distinguishable personality impressions.}
    \label{fig:prompt-validation}
\end{figure}

The live study did not include a direct participant-facing personality manipulation check after every phase, because doing so would have made the intended style framing obvious. Instead, behavioural trace measures and pre-study prompt validation were used to assess whether the assistant conditions were meaningfully distinct. Prompt validation indicated that the trait-linked prompts produced visibly different profiles, and the live trace data further showed differences in response length, user word share, and turn count across assistant conditions. The prompt validation profiles in Figure~\ref{fig:prompt-validation} indicate that the trait-linked prompts produced visibly different profiles; the live trace data further showed differences in response length, user word share, and turn count across assistant conditions.

\subsection{Ethical and Practical Controls}

The study used consent materials before participation and separated task instructions from system prompts so that the assistant did not receive hidden task-specific information unavailable to the participant. Participants were not asked to make real purchases, travel bookings, or health decisions. The health/diet task was framed as an information-seeking scenario rather than as medical advice, and the analysis treated trust and delegation in that condition as a user-experience outcome rather than as evidence of clinical correctness. These controls are important because personality manipulation can otherwise become confounded with persuasion, compliance, or perceived authority.

\section{Experiment Results}

\subsection{Assistant Conditions Were Behaviourally Distinct}

Before answering the research questions, we first checked whether the assistant conditions produced different interaction patterns in practice. Table~\ref{tab:behaviour} shows descriptive behavioural trace differences collapsed across the complete participant sample. Titan produced the longest assistant responses, with a mean of 146.38 words. Europa produced the shortest responses, with a mean of 90.22 words. Neptune was intermediate at 114.86 words.

A complementary pattern appeared for user word share. Europa had the highest user word share ($M=0.210$), followed by Neptune ($M=0.188$), while Titan had the lowest ($M=0.155$). Turn count showed a similar tendency: Europa produced the highest average number of turns per interaction ($M=4.85$), compared with Neptune ($M=4.31$) and Titan ($M=4.19$). These results suggest that the assistants differed not only by label but by live interaction structure. Titan was more expansive and assistant-dominant, Europa was more restrained and left more space for participant contribution, and Neptune occupied a middle position.

\begin{table}[t]
\centering
\caption{Behavioural trace descriptive statistics by assistant condition.}
\label{tab:behaviour}
\small
\begin{tabular}{lccc}
\toprule
\textbf{Assistant} & \textbf{Style} & \textbf{Assistant words} & \textbf{User share / turns} \\
\midrule
Titan & Extraverted & 146.38 & 0.155 / 4.19 \\
Europa & Conscientious & 90.22 & 0.210 / 4.85 \\
Neptune & Neutral & 114.86 & 0.188 / 4.31 \\
\bottomrule
\end{tabular}
\vspace{-0.2in}
\end{table}

\subsection{RQ1: Participant Personality and Information-Seeking Behaviour}

RQ1 examined whether participants' own personality traits directly predicted observable conversational information-seeking behaviour. In the analysis plan, participant Extraversion and Conscientiousness were treated as predictors of behavioural trace measures, including turn count, user word share, user message length, follow-up turns, user question count, and interaction duration. The corrected inferential analyses did not show robust direct effects after Holm correction \cite{holm79}. \textcolor{black}{The clearest directional signal was average user message length: more extraverted participants tended to write slightly longer messages, but this remained suggestive rather than statistically supported ($p=.085$). No clear effects were found for turn count, total user words, user word share, duration, follow-up turns, user question count, or user question share.} In other words, participant personality alone did not reliably predict whether participants produced more turns, contributed a larger share of words, asked more questions, followed up more often, or spent longer in the interaction.

\begin{figure}[t]
\centering
\includegraphics[width=\linewidth]{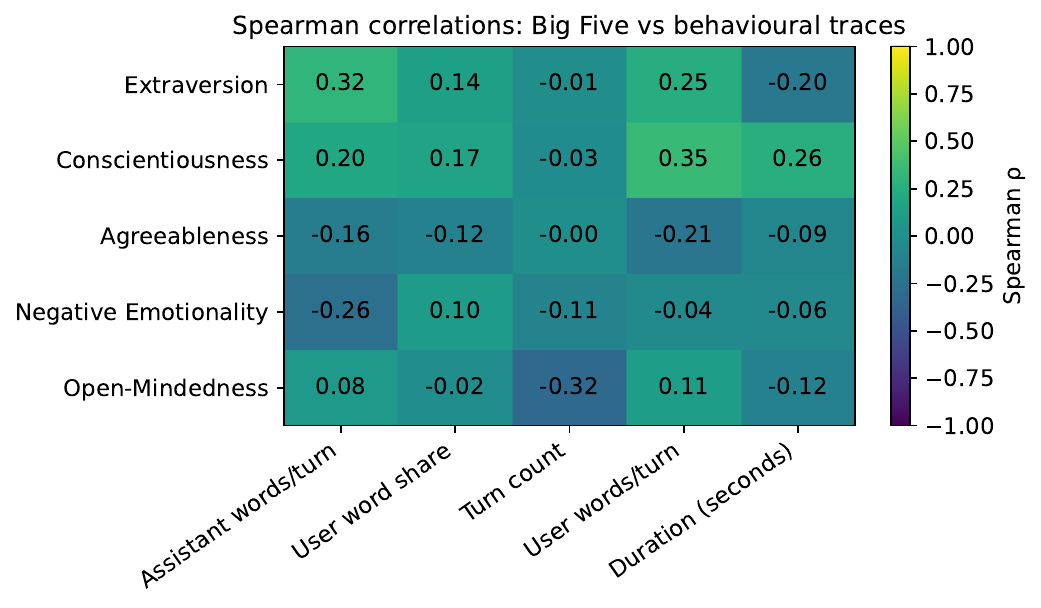}
\caption{Spearman correlations between Big Five traits and behavioural traces at the participant level.}
\vspace{-0.15in}
\label{fig:bfi_behavior_corr}
\end{figure}

Figure~\ref{fig:bfi_behavior_corr} provides a complementary descriptive view of this pattern by summarizing participant-level Spearman correlations between the Big Five traits and behavioural traces. The heatmap reinforces the same conclusion: the trait--behaviour relationships were scattered rather than systematic. Some isolated associations were visible, but they did not form a coherent behavioural profile. For example, Extraversion and Conscientiousness showed limited associations with message-length-related measures, while other traits showed similarly isolated patterns. However, these relationships were not consistent across behavioural traces and should therefore be interpreted descriptively rather than as evidence of stable trait-driven interaction behaviour.

The answer to RQ1 is therefore mostly negative. Broad participant traits did not translate into simple, uniform behavioural differences in conversational information seeking. This null result is substantively useful because it suggests that user personality should not be treated as a direct proxy for surface interaction behaviour. Rather than assuming that extraverted or conscientious users will consistently talk more, ask more, or interact for longer, the later analyses indicate that personality becomes more informative when considered as part of participant--assistant compatibility and task-specific style fit.

\subsection{RQ2: Personality Compatibility and Participant Differences}

RQ2 examined whether assistant personality effects depended on participant personality. Unlike RQ1, which tested direct trait--behaviour relationships, RQ2 focused on personality compatibility: whether participants evaluated an assistant more positively when its expressed style aligned with their own Extraversion--Conscientiousness profile.

The compatibility analysis focused on the two trait-linked assistant conditions, Titan and Europa. Compatibility was operationalised as the participant's relative trait balance, $z(\mathrm{Extraversion}) - z(\mathrm{Conscientiousness})$. Positive values indicated a more Extraversion-weighted profile, whereas negative values indicated a more Conscientiousness-weighted profile. This score was used to describe relative alignment between participant traits and assistant style, rather than to assign participants to fixed or pure personality types.

\begin{figure}[t]
    \centering
    \includegraphics[width=\linewidth]{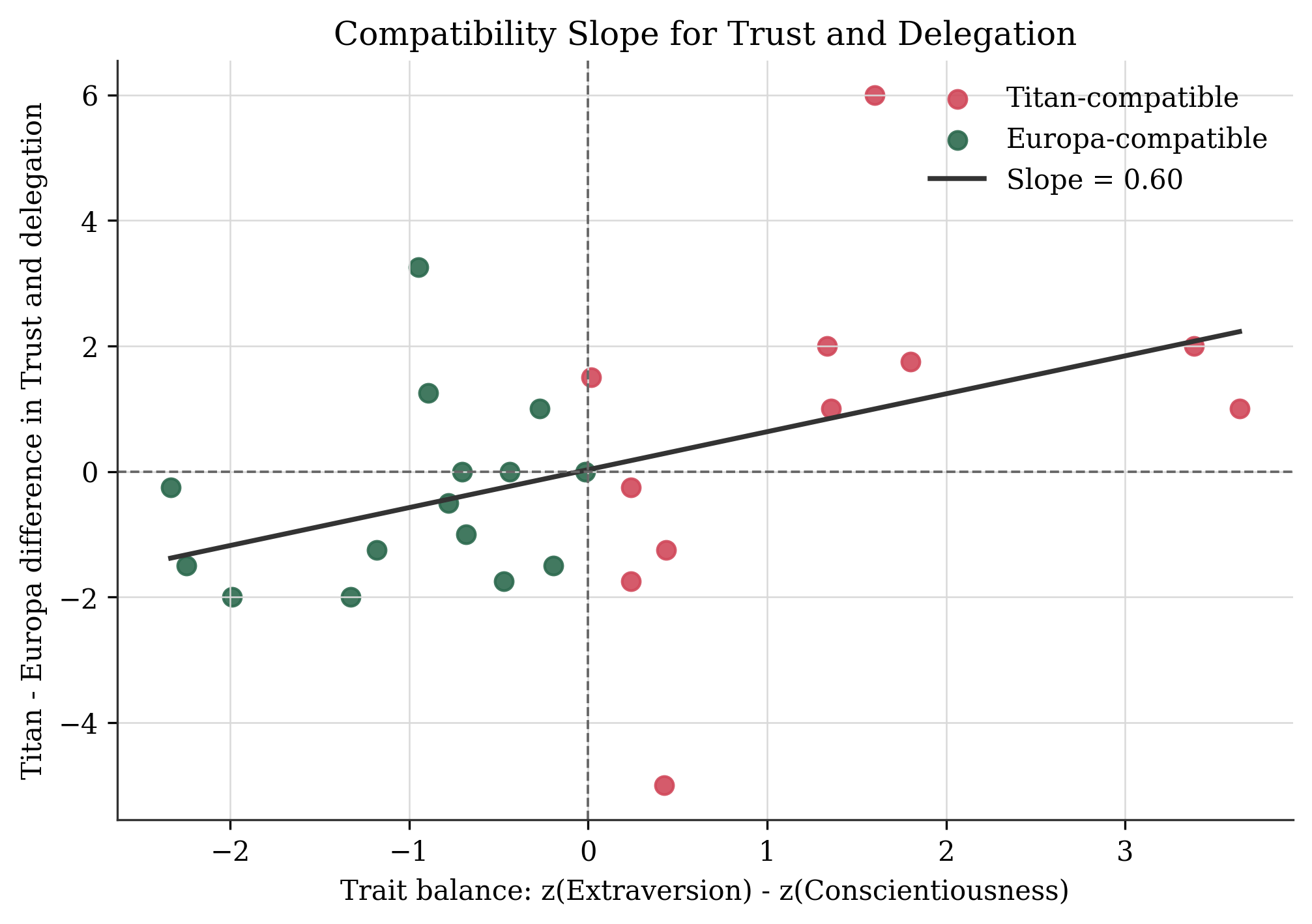}
    \caption{Compatibility slope for the Titan–Europa difference in trust and delegation. Positive values indicate participants who were relatively more extraverted than conscientious; negative values indicate the reverse. Titan tended to perform better relative to Europa as the Extraversion–Conscientiousness trait balance became more Extraversion-weighted.}
    \label{fig:compatibility-trust}
\end{figure}

The clearest compatibility pattern appeared for trust and delegation. \textcolor{black}{Inferentially, the Titan--Europa compatibility slope was significant for trust and delegation ($b=0.648$, $F(1,22)=6.38$, $p=.019$, partial $\eta^2=.225$). This indicates that, as participants became relatively more extraverted than conscientious, Titan tended to outperform Europa more strongly on trust and delegation.} Figure~\ref{fig:compatibility-trust} visualizes this relationship by plotting the Titan--Europa difference in trust and delegation against participants' Extraversion--Conscientiousness trait balance. Participants who were relatively more extraverted tended to evaluate Titan more favourably than Europa on trust and delegation, whereas participants who were relatively more conscientious tended to show the opposite tendency. This pattern suggests that trust and willingness to rely on the assistant were sensitive to whether the assistant's expressed style fit the participant's own interaction orientation.

The same compatibility direction was also visible for the overall post-interaction composite, but the pattern was weaker than for trust and delegation. \textcolor{black}{For the overall post-interaction composite, the compatibility effect was in the same direction but only directional rather than significant ($b=0.383$, $F(1,22)=2.98$, $p=.098$, partial $\eta^2=.119$). This supports interpreting compatibility as most clearly trust-related rather than as a broad improvement across all evaluation measures.} In other words, personality compatibility appeared most clearly in trust-related evaluations rather than in overall interaction quality. This distinction is important because it suggests that style fit may not uniformly improve all aspects of the experience. Instead, compatibility seems especially relevant to whether participants felt comfortable relying on the assistant.

\begin{figure*}[t]
\centering
\includegraphics[width=0.6\linewidth]{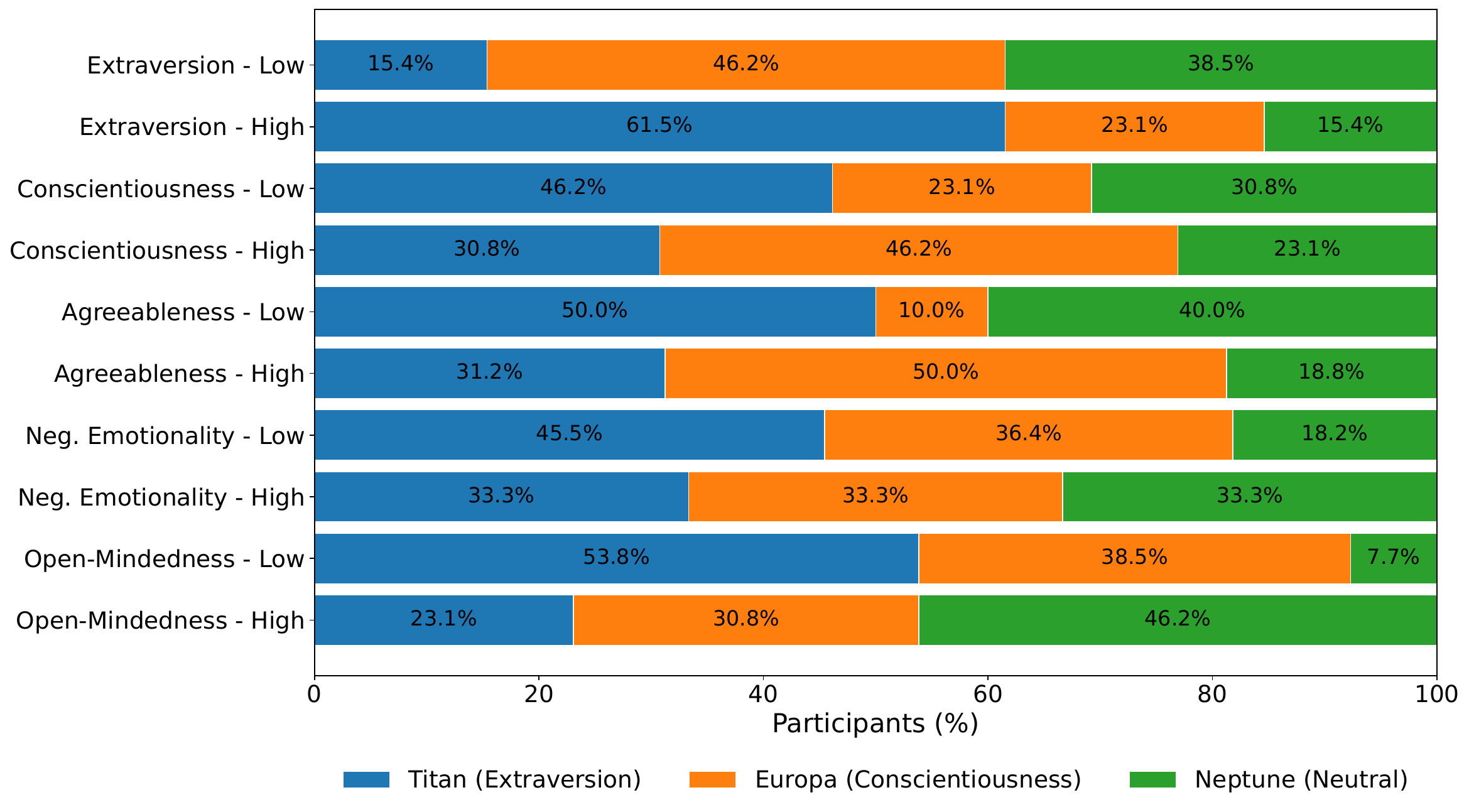}
\vspace{-0.15in}
\caption{Assistant preference distribution by median-split Big Five groups for the overall post-interaction composite. Bars show the percentage of participants whose highest overall composite score was assigned to Titan (Extraversion), Europa (Conscientiousness), or Neptune (Neutral).}
\label{fig:bfi_top_overall}
\end{figure*}

\begin{figure*}[t]
\centering
\includegraphics[width=0.6\linewidth]{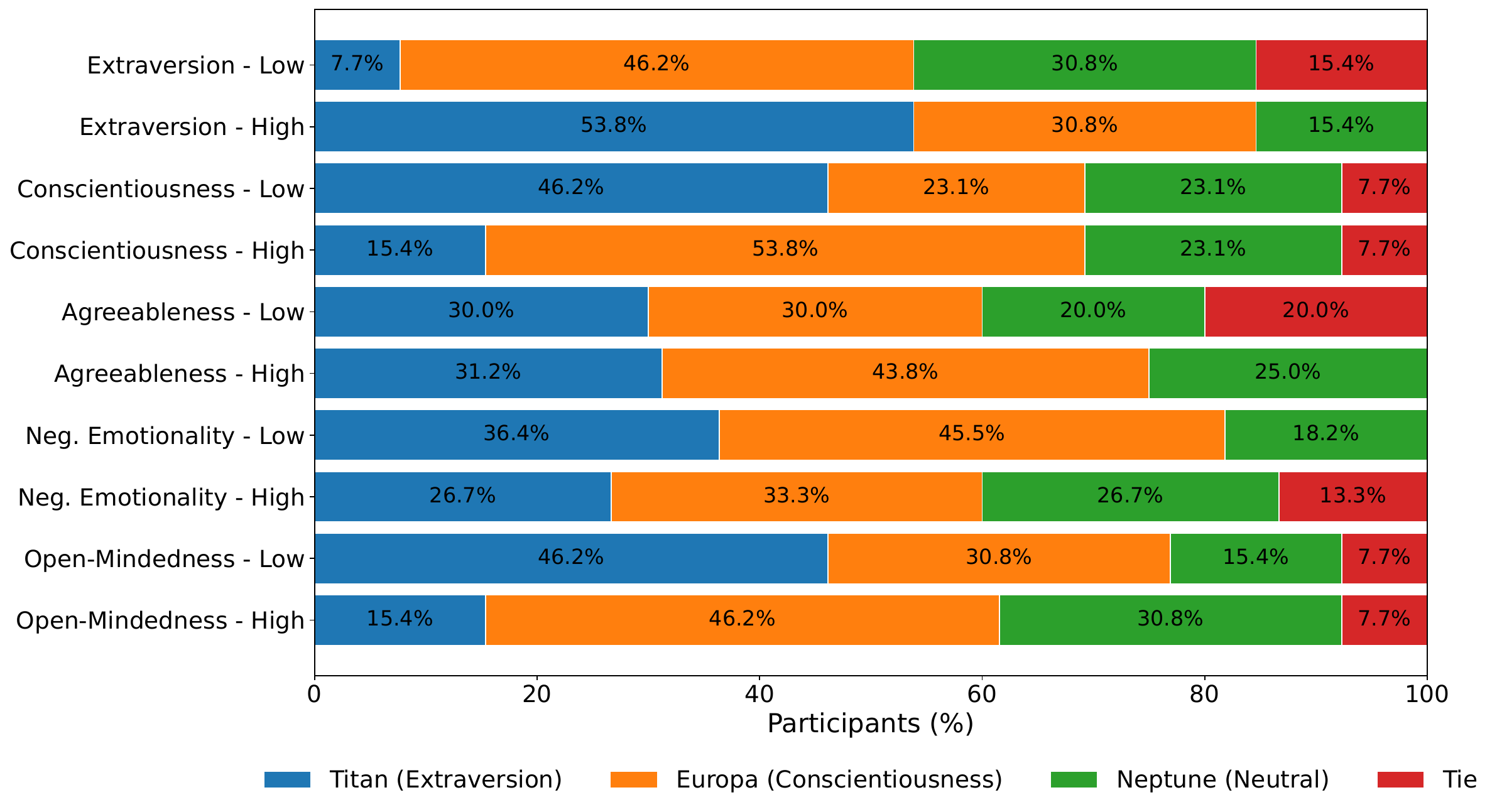}
\vspace{-0.15in}
\caption{Assistant preference distribution by median-split Big Five groups for trust and delegation. Bars show the percentage of participants whose highest trust and delegation rating was assigned to Titan (Extraversion), Europa (Conscientiousness), Neptune (Neutral), or a tie between assistants.}
\label{fig:bfi_top_trust}
\vspace{-0.15in}
\end{figure*}

Figures~\ref{fig:bfi_top_overall} and~\ref{fig:bfi_top_trust} provide a broader descriptive view by showing which assistant was top-rated within Big Five high/low groups. These figures include Neptune as the neutral baseline, whereas the compatibility analysis above focuses on the two trait-linked conditions. They should therefore be interpreted as descriptive triangulation rather than as additional inferential tests. The descriptive distributions were broadly consistent with the compatibility interpretation: participants higher in Extraversion more often favoured Titan, while participants higher in Conscientiousness more often favoured Europa, especially for trust and delegation.

Taken together, RQ2 indicates that participant personality mattered more as a moderator of assistant evaluation than as a direct predictor of conversational behaviour. The evidence does not support a deterministic matching rule, and participants with similar trait profiles did not always prefer the same assistant. Rather, the results suggest a compatibility tendency: assistant personality may be more effective when its expressed interaction style aligns with the participant's preferred way of engaging with the system, with the clearest implications for trust and delegation.

\subsection{RQ3: Assistant Personality Across Task Types}

RQ3 examined whether the effectiveness of assistant personality expressions varied across information-seeking task types. This question shifts the analysis from participant--assistant compatibility to task--assistant compatibility: whether the same expressed assistant style supports exploratory planning, comparative decision making, and verification-sensitive advice in the same way.

The clearest task-dependent pattern appeared for trust and delegation. As shown in Table~\ref{tab:taskmeans}, the assistant receiving the highest trust/delegation evaluation differed across tasks: \textcolor{black}{Titan was rated most favourably in Travel, Neptune in Shopping, and Europa in Health. Inferentially, this was the strongest supported effect in the study: the Task $\times$ Assistant interaction was significant for trust and delegation ($F(2,46)=3.55$, $p=.037$, partial $\eta^2=.134$).}

\textcolor{black}{The overall post-interaction composite followed the same descriptive ordering, although the inferential support was weaker and should be treated as directional rather than confirmatory ($F(2,46)=3.17$, $p=.051$). Titan was again most favourable in Travel, Neptune in Shopping, and Europa in Health. Because the omnibus interaction was statistically supported only for trust and delegation, whereas broader post-interaction measures were weaker, these results are better interpreted as task--style fit rather than as evidence for one globally superior assistant personality.}

\begin{table}[t]
\centering
\caption{Post-interaction means by task and assistant condition. Trust and delegation had the clearest supported Task $\times$ Assistant interaction; the overall post-interaction composite followed the same descriptive pattern but had weaker statistical support.}
\label{tab:taskmeans}
\small
\setlength{\tabcolsep}{1pt}
\renewcommand{\arraystretch}{0.8}
\begin{tabular}{lcccccc}
\toprule
\multirow{2}{*}{\textbf{Task}} & \multicolumn{3}{c}{\textbf{Trust / delegation}} & \multicolumn{3}{c}{\textbf{Overall post-interaction composite}} \\
\cmidrule(lr){2-4}\cmidrule(lr){5-7}
 & \textbf{Titan} & \textbf{Europa} & \textbf{Neptune} & \textbf{Titan} & \textbf{Europa} & \textbf{Neptune} \\
\midrule
Travel & 5.62 & 5.03 & 4.50 & 5.83 & 5.08 & 4.77 \\
Shopping & 4.72 & 4.97 & 5.94 & 5.22 & 5.12 & 5.80 \\
Health & 5.12 & 5.53 & 4.58 & 5.27 & 5.59 & 4.94 \\
\bottomrule
\end{tabular}
\vspace{-0.15in}
\end{table}
To understand why the preferred assistant style changed across tasks, we coded non-empty post-interaction comments into recurring themes. Table~\ref{tab:qualitative_themes} summarizes these themes and provides qualitative support for the quantitative Task $\times$ Assistant pattern.

\begin{table*}[t]
\centering
\caption{Interaction themes from lightweight coding of non-empty post-interaction comments. Comments could receive multiple codes; excerpts are anonymised by participant ID.}
\renewcommand{\arraystretch}{0.8}
\vspace{-0.15in}
\label{tab:qualitative_themes}
\begin{tabular}{p{3.2cm} c p{5.0cm} p{5.6cm}}
\toprule
\textbf{Interaction theme} & \textbf{Mentions} & \textbf{Interpretation} & \textbf{Example excerpt} \\
\midrule
verbosity/directness mismatch & 7 & Users rejected styles that felt too chatty, emotional, fluffy, or insufficiently direct for the task. & GU26, Shopping/Titan: “For a technical task, I expected more technical dialogue, less emotions.” \\
constraint adherence / specificity & 6 & Several comments judged whether the assistant maintained constraints, surfaced alternatives, or gave concrete enough recommendations. & GU23, Shopping/Europa: “it failed to showcase potentially better alternatives and focused on famous, brand items” \\
proactive follow-up / momentum & 5 & Follow-up questions were valued when they advanced exploration and reduced the user's planning burden. & GU10, Travel/Titan: “relevant follow up questions … closely aligned with my next task” \\
health trust/evidence grounding & 5 & Verification-sensitive tasks prompted concerns about evidence, research links, independent checking, and over-agreement. & GU01, Health/Neptune: “linking literature or research would be more beneficial” \\
excessive questioning / slow convergence & 4 & The same follow-up behaviour became negative when users wanted quick recommendations or forward progress. & GU18, Shopping/Europa: “kept asking me questions when I wanted some quick recommendations” \\
accuracy/evidence/detail concerns & 4 & Participants flagged price, alternative options, product detail, or depth of scientific explanation as important. & GU01, Shopping/Europa: “recommended s23 to be in my budget … but a quick google search showed” \\
balanced low-friction style & 2 & Neutral or shorter styles were praised when they felt less overwhelming and avoided over-questioning. & GU10, Health/Neptune: “kept follow up questions to a minimum not overloading my brain” \\
\bottomrule
\end{tabular}
\vspace{0.35em}
\begin{minipage}{0.98\linewidth}
\footnotesize\textit{Note.} Themes were used interpretively to contextualize quantitative interaction patterns, not to establish standalone qualitative claims.
\end{minipage}
\end{table*}

The qualitative themes clarify why the same conversational behaviour could be interpreted differently across task contexts. In Travel, proactive follow-up and conversational momentum were useful because the task required preference disclosure, option exploration, and iterative refinement. This helps explain why Titan's energetic and collaborative style fit the exploratory travel-planning task: its follow-up questions appeared to make the next step easier for participants and helped the interaction continue.

In Shopping, however, participants placed greater emphasis on directness, specificity, and efficient convergence. The themes of verbosity/directness mismatch, constraint adherence, and excessive questioning indicate that a style perceived as lively or methodical could become costly when users wanted quick, concrete recommendations. In this context, repeated follow-up questions or affective language could be experienced as friction rather than support. This helps explain why Neptune, the neutral and lower-friction baseline, performed best descriptively in the comparative shopping task.

In Health, the central criterion shifted again. Participants were less concerned with conversational momentum and more concerned with evidence, caution, and verifiability. The health trust/evidence grounding theme shows that users wanted claims to be supported by research, links, or independent checking, while the accuracy/evidence/detail theme shows concern about whether the assistant provided enough reliable detail. This supports the interpretation that Europa's more structured and careful style was better aligned with verification-sensitive information seeking.

The answer to RQ3 is therefore positive but task-specific. Assistant personality effects differed across task types, but not because one assistant was globally superior. Instead, the results show a task--style fit pattern: Titan appeared strongest for exploratory Travel, Neptune for comparative Shopping, and Europa for verification-sensitive Health. The pattern was clearest for trust and delegation, while the overall post-interaction composite followed the same direction more weakly. Together, the quantitative ordering and qualitative themes suggest that assistant personality should be treated as a situational interaction design variable rather than as a universal default setting.

\subsection{Summary of Findings}

Across the results, the main finding is that assistant personality mattered, but in a contextual rather than universal way. The three assistant conditions were behaviourally distinguishable, indicating that prompt-defined personality styles changed the live interaction structure even when the base model and task materials were held constant. However, participant personality alone did not robustly predict behavioural traces such as turn count, question behaviour, user word share, or message length. This suggests that broad personality traits should not be treated as simple predictors of how users will behave in conversational information seeking. Instead, personality became more informative when analysed relationally: participant--assistant compatibility was associated with trust and delegation, especially along the Extraversion--Conscientiousness contrast that was directly expressed in the assistant prompts.

Assistant effects also varied by task type, showing that style fit depended on the information-seeking ecology. Titan was strongest for exploratory travel planning, where proactive and socially engaging interaction could support preference articulation and conversational momentum. Neptune performed best in comparative smartphone shopping, where a lower-friction style appeared better aligned with efficient narrowing and concrete decision support. Europa was strongest for verification-sensitive health/diet information seeking, where structured and cautious interaction better supported trust and delegation. \textcolor{black}{The exit-reveal results further reinforced this interpretation. Learning the style framing did not substantially change preferences: responses to ``Knowing this changes which system I prefer'' were significantly below the midpoint ($M=1.54$, $W=7.0$, $p<.001$). By contrast, participants rated both explicit style choice ($M=6.19$, $W=11.0$, $p<.001$) and automatic style adaptation ($M=6.08$, $W=26.5$, $p<.001$) significantly above the midpoint.} Taken together, these findings suggest that assistant personality should be treated as a task- and user-sensitive interaction design variable rather than as a single globally optimal persona.

\section{Conclusion}

This paper examined assistant personality as a controllable interaction variable in conversational information seeking. In a within-subject study with 26 participants, three assistant styles, and three task ecologies, the results did not support a single universally best persona. Instead, the clearest pattern was task--style fit: Titan was strongest for exploratory travel planning, Neptune for comparative shopping, and Europa for verification-sensitive health information, with trust and delegation showing the clearest supported task-by-assistant interaction. Participant personality was not a strong direct predictor of surface trace behaviour, but compatibility between participant and assistant style was associated with trust-related outcomes. This distinction is important for personality-aware conversational search: personality should not be reduced to a user trait, an assistant label, or a verbosity setting. It is expressed through pacing, initiative, evidence, and structure, and it is interpreted in relation to the user's task.

These findings suggest that assistant personality should be treated as a configurable and task-sensitive interaction parameter rather than as a fixed default persona. For conversational search system design, this means that assistant style should be evaluated with task-specific outcomes, trust and delegation should be distinguished from general liking, and personalization should combine user control with transparent adaptation rather than encouraging unwarranted reliance. Future work should combine personality-aware interaction analysis with stronger answer-quality evaluation, grounding checks, and larger, more diverse samples. Future work could further explore richer multimodal interaction and modeling \cite{fu2026stream,ye2026multimodal,zheng2026we,zhuang2025frequency,he2025double}. Follow-up studies should also incorporate semi-structured interviews to examine why participants perceived particular styles as better suited to specific tasks, whether style shaped decision confidence, and which behaviours increased trust, pressure, or friction. Such interview data would enable fuller qualitative coding and thematic analysis of the perceptions and mechanisms underlying the observed interaction patterns.  The present study nevertheless suggests that conversational search systems should treat assistant style as part of the information-seeking interface itself.

\section{Limitations}

Several limitations should be considered when interpreting the findings. First, the study used a modest convenience sample, so the results should be viewed as evidence of plausible interaction patterns rather than population-level estimates. The within-subject design, however, helped reduce between-participant variability and supported the identification of consistent task- and style-related tendencies.

Second, assistant personality was implemented through prompting rather than model training. The manipulation therefore reflects expressed conversational style rather than a stable model trait, which aligns with how many current LLM systems expose personality-like behaviour to users.

Third, the study examined only two Big Five dimensions, Extraversion and Conscientiousness, across three information-seeking task types. Other traits, domains, or higher-stakes settings may yield different patterns.

Finally, the analysis focused on interaction behaviour and user experience rather than exhaustive factual verification. Measures such as response length, turn count, and question share describe how interactions unfolded, but not the full semantic quality of the information exchanged. Future work should combine personality-aware interaction analysis with stronger answer-quality evaluation, grounding checks, and larger, more diverse samples.

\section{GenAI Usage Disclosure}
The authors used ChatGPT to support language refinement and writing. All AI-assisted text was reviewed and edited by the authors, who take full responsibility for the paper’s content. Generative AI tools were not used to produce experimental results, fabricate data, or generate citations.

\balance
\bibliographystyle{ACM-Reference-Format}
\bibliography{references}

\end{document}